\documentclass{aa}  

\usepackage{longtable}
\usepackage{graphicx}
\usepackage{txfonts}
\usepackage{longtable,lscape}
\usepackage{booktabs} 
\def\R23{\mbox{$\rm R_{23}$}}

\def\kmsmpc{km s$^{-1}$ Mpc$^{-1}$}

\def\Hb{\mbox{${\rm H}{\beta}$}}
\def\Ha{\mbox{${\rm H}{\alpha}$}}
\def\OIIIa{\mbox{${\rm [O\,III]\,}{\lambda\,5007}$}}

\def\OII{\mbox{${\rm [O\,II]\,}{\lambda\,3727}$}}

\def\NII{\mbox{${\rm [N\,II]\,}{\lambda\,6584}$}}

\begin{document}

%\title{Slow-then-rapid quenching  as traced by enhanced metallicities of LoCuSS cluster galaxies in the slow quenching phase}
\title{Slow-then-rapid quenching as traced by tentative evidence for enhanced metallicities of cluster galaxies at $z \sim 0.2$ in the slow quenching phase}

%\and

\author{C.~Maier\inst{1}
\and B.\,L.~Ziegler\inst{1}
\and C.\,P.\,~Haines\inst{2}
\and G.\,P.\,~Smith\inst{3}
}

\institute{University of Vienna, Department of Astrophysics, Tuerkenschanzstrasse 17, 1180 Vienna, Austria\\
\email{christian.maier@univie.ac.at}
%2
\and INAF - Osservatorio Astronomico di Brera, via Brera 28, 20121 Milano, Italy
\and School of Physics and Astronomy, University of Birmingham, Edgbaston, Birmingham, B15, 2TT, UK
}

\titlerunning{Quenching, metals and star formation in cluster galaxies}
\authorrunning{C. Maier et al.}

%##################################################################################
\date{Received ; accepted}

\abstract 
{}
{%Aims
As large scale structures in the Universe develop with time, environmental effects become more and more important as a star-formation quenching mechanism. Since the effects of environmental quenching are more pronounced in denser structures which form at later times,
we want to constrain environmental quenching processes using cluster galaxies at $z<0.3$.
}
%Methods
{
We explore seven  clusters from the Local Cluster Substructure Survey (LoCuSS) at $0.15<z<0.26$ with spectra of 1965
%about 2700 
cluster members in a mass-complete sample
from the ACReS  Hectospec survey  covering a region 
 which corresponds to about three virial radii for each cluster. We measure fluxes
of \OII, \Hb, \OIIIa, \Ha, and \NII\, emission lines  of cluster members enabling us to unambiguously derive O/H gas metallicities, and also star formation rates from extinction-corrected \Ha\, fluxes. We compare our cluster galaxy sample with a field sample of 705
%about 1000 
galaxies at similar redshifts observed with Hectospec as part of the same survey.
}
%Results
{
We find that star-forming cluster and field galaxies show similar median specific SFRs in a given mass bin ($1-3.2 \cdot 10^{10}\rm{M}_{\odot}$ and $3.2-10 \cdot 10^{10}\rm{M}_{\odot}$, respectively), 
but their O/H values are displaced, in the lower mass bin, to higher values (significance $2.4 \sigma$) at projected radii of $R<R_{200}$ compared with galaxies at larger radii and in the field.
%for the inner regions of the clusters ($R<R_{200}$) 
%compared to field and $R>R_{200}$ galaxies in the lower mass bin.
The comparison with metallicity-SFR-mass model predictions with inflowing gas indicates a slow-quenching scenario in which strangulation is initiated when galaxies pass $R \sim R_{200}$ by  stopping the  inflow of 
gas. We find tentative evidence that the metallicities of cluster members inside  $R_{200}$ are thereby increasing, but their  SFRs are hardly affected for a period of time, 
because these galaxies consume available disk gas.
We use the observed fraction of star-forming cluster galaxies as a function of clustercentric radius compared to predictions from the Millennium simulation to constrain quenching timescales to be  1$-$2\,Gyrs, defined as the 
time between the moment the galaxy passes $R_{200}$ until complete quenching of star formation. 
This is consistent with a slow-then-rapid quenching scenario. Slow quenching (strangulation) starts when the gas inflow is stopped when the galaxy passes $R_{200}$ with a %``delayed'' 
phase in which cluster galaxies are still star-forming,
 but they show elevated metallicities tracing the ongoing quenching.  This phase
lasts for 1$-$2\,Gyrs, meanwhile the galaxies travel to denser inner regions of the cluster, and is followed by a  ``rapid'' phase: a rapid complete quenching of star formation due to the increasing ram-pressure towards the cluster center which can also strip the cold gas in massive galaxies.
}{}

%A maximum of 6 key words should be listed after the abstract. These must be selected from a list that is published each year in the first issue in January.

\maketitle

%#######################################################################

%\newpage

\setcounter{section}{0}
%##############################################################################
\section{Introduction}
%#########################################################
%%%%%%%%%%%%%%%%%%%%%%

~~~ The internal properties of galaxies in dense environments are known to differ from isolated galaxies, as seen in their color \citep[e.g.,][]{peng10}, star formation rate \citep[SFR; e.g.,][]{wetzel12} or morphology \cite[e.g.,][]{dressler80}. However, a detailed understanding of the \emph{physics} responsible for the differences between field and cluster galaxies has so far proved difficult, although a number of mechanisms have been proposed that could play a role in affecting and stopping (quenching) star formation in dense environments.
%, including ram pressure stripping \citep[e.g.,][]{gunngott72}. 
A plausible cause of the quenching of star formation in galaxies after accretion into massive halos are their interactions with the dense intracluster medium (ICM) due to their high velocity passage through the dense ICM.

As a scenario for quenching satellite galaxies in groups and clusters, \citet{wetzel13} proposed a two-stage model in which star formation truncation happens rapidly, but only after a delay time after accretion into the cluster. In this model, galaxies experience little or no change in their SFRs during the delay phase. The model has been used to successfully interpret a number of observations at $z<1$, but the hypothesis that galaxies are completely unaffected by environment during the delay period remains controversial and has been a theoretical challenge. 
It is difficult to understand and to simulate galaxies travelling at $1000-2000$\,km/s through the cluster ICM and remaining uninfluenced by their environment for several Gyrs, and then suddenly stopping forming stars on a short timescale. 
On the other hand, \citet{haines13} found the  specific SFRs (SSFRs) of massive $\rm{log(M/M_{\odot})} > 10$ star-forming cluster galaxies within $R_{200}$ to be (slightly) lower than the SSFRs of their counterparts in the field, inferring that star-forming galaxies are slowly quenched upon accretion into clusters, with a  best fit by models in which their SFRs decline exponentially on quenching timescales in the range $0.7-2$\,Gyrs.  

%#########################################################
%
~~~Several authors have studied the role of the SFR  and of the specific SFR (SSFR$=$SFR/M$_{*}$) on the metallicity of a galaxy and presented links between SFR and metallicity \citep[e.g.,][]{elli08,lopez10}, finally claiming an epoch-independent fundamental metallicity relation (FMR) between metallicity, mass and SFR, Z(M$_{*}$,SFR), expected to be applicable at all redshifts  \citep[][]{mannu10}.
In order to explain star formation as a second parameter in the mass-metallicity relation (MZR), some publications have ascribed an 
%ad-hoc 
inflow of gas as the responsible driver for the  dilution of metallicity and increase of SFR \citep{mannu10,dave12,dayal13}. One explanation for the origin of this relation has been summarized
by \citet{lilly13} and is known as the bathtub model. The SFR is thereby
closely linked to the mass of gas in the galaxy modulated by infalling gas and outflows.
%

%%%%%%%%%%%%%%%%%%%%%%%%%%%%%%%%%%%%%%%%%%%%

~~~ \citet{peng15} used the SDSS local sample and found that the stellar metallicity of satellite galaxies is slightly higher than that of central galaxies. They inferred that lower mass satellite galaxies are more prone to strangulation, implying that environmental effects
are responsible for stopping the inflow of gas.
In \citet[][MKZ16 in the following]{maier16} we explored the Z(M$_{*}$,SFR) in the MACSJ0416.1-2403 cluster at $z \sim 0.4$ with a sample of mostly intermediate and low mass galaxies ($8.5<\rm{logM/M_{\odot}} < 10.2$), and only a handful of higher mass objects.
We found a dependence of metallicity on enviroment at a given stellar mass, manifested in higher metallicities in over-dense regions, which we interpreted as a strangulation scenario. 
We call strangulation/starvation a scenario in which the supply of cold gas (through cooling) onto the galaxy disk is halted because the \emph{hot} halo gas is stripped owing to external forces.
In a context of a rich cluster, this scenario was first described by \citet{larson80}, who used the removal of the gas reservoir due to external forces to explain the passive cluster galaxy population.
For infalling cluster galaxies, the gas reservoir may be stripped or truncated due to interactions with cluster potential, ICM gas, and/or other galaxies \citep[e.g.,][]{bekki02}. This can terminate the fresh pristine gas accretion on to the galaxy (which would otherwise dilute the interstellar medium), increasing the retained gas metallicity before the galaxy will completely stop forming stars, i.e., be completely ``quenched'', at a later time. Thus, the gas-phase metallicity can be used as a \emph{slow quenching} (strangulation) \emph{tracer} (cf. MKZ16).

In this paper we  extend these investigations to a larger sample of cluster galaxies,
measuring O/H abundances in K-band selected galaxy samples in seven massive clusters at $0.15<z<0.26$ 
%$z \sim 0.2$ 
from LoCuSS, 
and contrasting them to comparable measurements of field galaxies at similar redshifts observed at the same time and with the same Hectospec instrument configuration as the cluster galaxies.
 We do this using the two connected relations, the MZR and the mass$-$(S)SFR relation, studying the environmental dependence of these relations and the implications for the quenching processes.

~~~To estimate the chemical abundances, a number of 
diagnostics have been developed based on strong emission lines (ELs), 
 ${\rm [O\,II]\,}{\lambda\,3727}$, H$\beta$, ${\rm [O\,III]\,}{\lambda\,5007}$, H$\alpha$ and [NII]${\lambda\,6584}$.
At higher redshifts, these ELs move to the near-infrared, with H$\alpha$ and [NII]${\lambda\,6584}$ already shifting beyond optical above redshifts of $\sim 0.5$, requiring near-infrared spectroscopy to observe these ELs \citep[e.g.,][]{maier14,maier15}.
Thus, the $0.15<z<0.26$ epoch we probe is a redshift range at which all these ELs can be observed free of strong night skylines and 
with the same optical spectrograph (in our case Hectospec),
without the complication of obtaining a good relative calibration between spectra from optical and near-infrared spectrographs.

%#####################

%
In MKZ16 we studied star-forming galaxies in one cluster and found indication that
accreted cluster star-forming galaxies have similar SFRs to their field counterparts, but their gas-phase metallicities are enhanced due to an ongoing strangulation (slow quenching) process. 
In this study we reinforce the findings of MKZ16 by now studying seven clusters, and further extend the discussion of a plausible strangulation scenario, now in a \emph{slow-then-rapid} framework: stopping the inflow of gas hardly affects the SFRs of cluster galaxies during a \emph{slow-quenching} phase, but enhances the gas-phase metalicities of cluster galaxies compared to field counterparts, before the star formation is (rapidly) completely quenched at a later time.

%########################

  The paper is structured as follows: In Sect. 2 we
present the selection of the cluster EL galaxies at $z \sim 0.2$ and their Hectospec spectroscopy.
We investigate the AGN (active galactic nucleus) contribution
and present the derivation of SFRs, metallicities, and stellar masses of the cluster and field galaxies we observed.
In Sect.\,3 we present  the phase-space diagram, 
the MZR and the mass$-$SSFR relations at $z\sim 0.2$, and investigate their dependence on environment, also as a function of the clustercentric radius of the clusters. 
We discuss how the cluster environment affects the chemical enrichment. 
In Sect.\,4 we compare our observed mass-metallicity-SFR values with bathtub model predictions with inflow of pristine gas, finding indication for a strangulation scenario similar to the findings of MKZ16: the inflow of pristine gas is halted and strangulation is initiated when galaxies are accreted inside $R_{200}$ of the cluster.
We compare the position of galaxies in the phase-space diagram with theoretical predictions in order to constrain the time between infall into the cluster and complete quenching of star formation (quenching timescales).
We thereby compare the observed radial population gradients of star-forming galaxies with predictions from the Millennium simulation \citep{springel05}.
We then discuss the slow-then-rapid quenching scenario implied by our findings.
Finally in Sect.\,\ref{sec:summary} we summarize our conclusions.
A {\sl concordance}-cosmology with $\rm{H}_{0}=70$ \kmsmpc,
$\Omega_{0}=0.25$, $\Omega_{\Lambda}=0.75$ is used throughout this
paper.  
We assume a Salpeter \citep{salp55} initial mass function (IMF) for all derived stellar masses and SFRs  and correct existing measurements used in this paper to a Salpeter IMF. 
Note that {\sl
  metallicity} and {\sl abundance} will be taken to denote {\it oxygen abundance}, O/H, throughout this paper, unless otherwise specified.
In addition, we use dex throughout to denote the anti-logarithm,
i.e., 0.3\,dex is a factor of two.
%########################################################################################

\section{The data and measurements}

%#####################
\subsection{Cluster galaxies from LoCuSS and sample selections}
%########################################################################
%\begin{multicols}{3} 
\begin{figure*}[ht]

%WHAN7ClustSEP18.ps
\includegraphics[width=10cm,angle=270,clip=true]{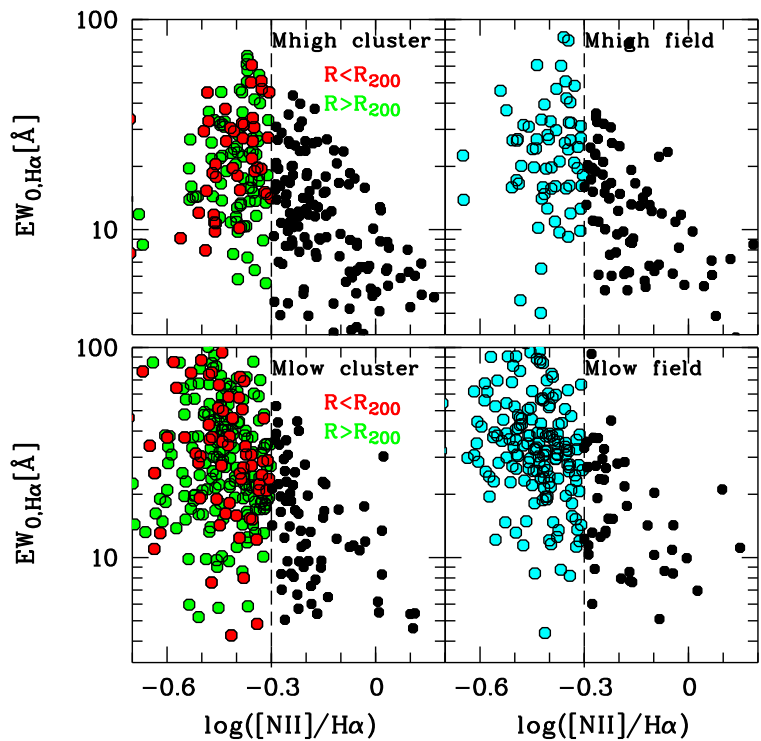}
\caption
{
\label{fig:WHAN} 
\footnotesize 
WHAN \citep{cidfern11} diagnostic diagram to distinguish star formation-dominated galaxies
from AGNs for the LoCuSS \emph{mass-complete} cluster and field galaxy samples.
410 cluster galaxies and 227 field galaxies are in the star-forming region of the WHAN diagram ($SFgals$ sample, red, green and cyan points), having a restframe $EW(H\alpha) >3$ \AA\, and log([NII]/H$\alpha)\leq -0.3$.
The red symbols show the subsample of $SFgals$ with $R<R_{200}$. 
Mlow (Mhigh) galaxies shown as black symbols in the lower (upper) pannels lie in the region occupied by Type-2 AGNs  ($SeyfLiners$ sample, see Tab.\,\ref{tab:subsamples}).
}
\end{figure*}
%\end{multicols}
%########################################################################################

%########################################################################

\begin{figure*}[ht]

\includegraphics[width=10cm,angle=270,clip=true]{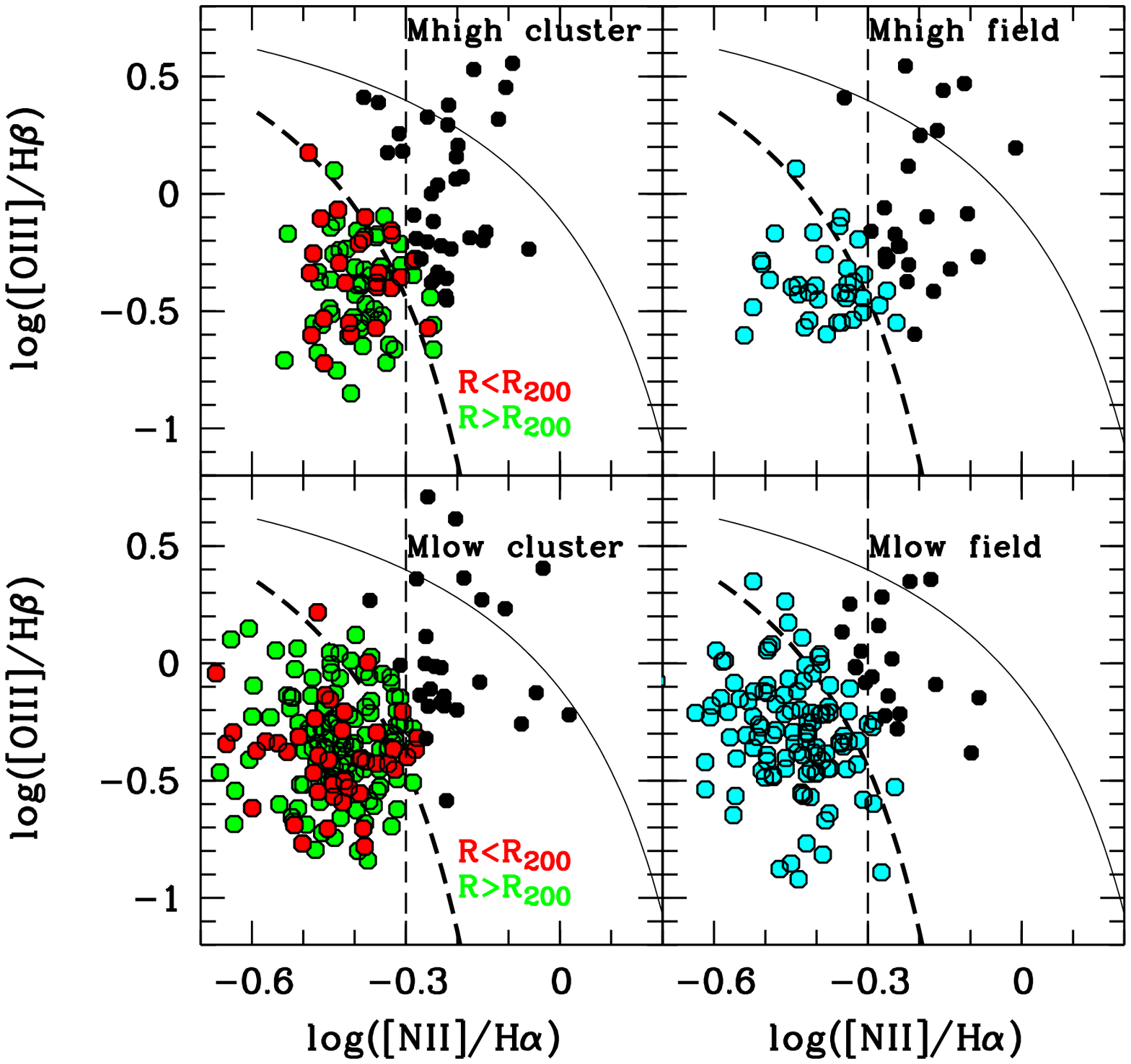}
\caption
{
\label{fig:BPT} 
\footnotesize 
BPT \citep{bald81} diagnostic diagram to distinguish star formation-dominated galaxies
from AGNs for the  \emph{mass-complete} cluster and field galaxy samples.
259 cluster galaxies and 169 field galaxies are in the star-forming region of the BPT diagram ($OHgals$ sample, red, green and cyan points), lying under and to  the left of the empirical (dashed) curve of \citet{kaufm03}. 
We also include in the  $OHgals$ sample a few objects above and up to a maximum of 0.3\,dex to the right of the \citet{kaufm03} curve, to take error bars of the flux measurements into account, and to mimimize the number of excluded composite galaxies above but close to the demarcation line which are likely to have their H$\alpha$ line mostly produced by star-formation. 
The red symbols show the subsample of $OHgals$ with $R<R_{200}$. 
Mlow (Mhigh) galaxies shown as black symbols in the lower (upper) pannels lie in the region occupied by AGN and composite galaxies ($CompAGNs$ sample, see Tab.\,\ref{tab:subsamples}), with composites lying between the  empirical dashed curve of \citet{kaufm03} and the theoretical solid curve of \citet{kewley01}.
The vertical dashed line at log([NII]//H$\alpha$)$=-0.3$ shows the demarcation line for selecting $SeyfLiners$ galaxies (log([NII]/H$\alpha$)$>-0.3$) using the criterium of \citet{cidfern11}, see Fig.\,\ref{fig:WHAN}.
%
%The histograms on the right show that Mlow galaxies of the $OHgals$ sample at $R<R_{200}$ (red) have in general lower [OIII]/H$\beta$ ratios than cluster galaxies at $R>R_{200}$ (green) and field galaxies (cyan), a sign of being in the act of quenching star formation (see more details in Sect.\,\ref{sect:MZR}).
%
}
\end{figure*}

%########################################################################################

The LoCuSS (Local Cluster Substructure Survey) is a multi-wavelength survey of X-ray luminous galaxy clusters at $0.15<z<0.3$ \citep{smith10} drawn from the ROSAT All Sky Survey cluster catalog. These clusters benefit from a rich data set, including: Subaru/Suprime-Cam optical imaging, Spitzer/MIPS 24\,$\mu$m maps, Herschel/PACS$+$SPIRE 100-500\,$\mu$m maps, Chandra and XMM X-ray data, GALEX UV data, and near-infrared imaging. All of this data cover \emph{at least} 25$\times$25 square arcminutes fields centered on the clusters.

The ACReS (Arizona Cluster Redshift Survey) is a large spectroscopic survey using Hectospec/MMT of 30 massive clusters of LoCuSS, seven out of which we study in this paper (see Table\,\ref{tab:7Clusters}). The 270 line grating was used, providing a wide wavelength range ($3650-9200$\AA) at 6.2\AA\ resolution.
The main strengths of ACReS are its wide field-of-view (FoV$\sim 1$ degree diameter), which reaches well into the infall regions (up to about three virial radii) of the clusters, and the careful target selection based on J and K near-infrared imaging, which provides an unbiased, mass-complete sample of cluster galaxies, down to  $\rm{log(M/M_{\odot})} \sim 10$ \citep[cf.][]{haines15}.
We note that, while the stellar masses derived by \citet{haines15} used an older relation of \citet{bell03}, we now use a newer relation from \citet{zibet09} to derive stellar masses, as described in Sect.\,\ref{Oxabund}, and get a slightly different mass-completness limit of $\rm{log(M/M_{\odot})} \sim 10$.
%of $\rm{log(M/M_{\odot})} \ge 10$ \citep[compared to $\rm{log(M/M_{\odot})} \ge 10.3$ of][]{haines15}.
%

%###########################################################################################

%Tabelle  Clusters
\begin{table*}
\caption{The seven clusters of this study: mean redshift of cluster members, coordinates, radii $R_{200}$ and cluster masses $M_{200}$ \citep[as reported in][]{haines13}; number of cluster members, star-forming (SF) objects and galaxies with O/H measured in our sample. The '200' subscript refers to the radius where
the average density is 200 times the critical density of the Universe at the cluster redshift.}
\label{tab:7Clusters}
\begin{tabular}{ccccccccc}
\hline\hline      
Cluster & z &   RA    &    DEC & $R_{200}$ & $M_{200}$             &  cluster                     &   SF cluster   &  members  \\
Name    &   &         &        & (Mpc)    &  ($10^{14}M_{\odot}$) & members\tablefootmark{a}    &   members\tablefootmark{a}      &  with O/H\tablefootmark{a} \\ 
\hline      
RXJ1720& 0.1599 & 17:20:10.14  & +26:37:30.90  & 2.02 & 9.40  &  358/219 &   63/22   &  45/15  \\ %& 9.5  \\
A1914 & 0.1671 & 14:25:59.78  & +37:49:29.10  & 1.90 & 7.69  &  388/276 &   101/50  &  65/31  \\ %& 9.5  \\
A1689 & 0.1851 & 13:11:29.45  & -01:20:28.30  & 2.13 & 11.01 &  438/291 &   116/46  &  66/32  \\ %& 9.65 \\
A963  & 0.2043 & 10:17:01.20  & +39:01:44.40  & 1.81 & 6.75  &  414/332 &   107/75  &  69/45  \\ %& 9.75 \\
A2390 & 0.2291 & 21:53:36.72  & +17:41:31.20  & 2.21 & 12.36 &  371/287 &   86/70   &  61/48  \\ %& 10   \\
A1763 & 0.2323 & 13:35:16.32   & +40:59:45.60  & 1.85 & 7.18  &  292/230 &   76/63   &  53/43  \\ %& 10   \\
A1835 & 0.2520 & 14:01:02.40  & +02:52:55.20  & 2.27 & 13.34 &  465/330 &   112/84  &  62/45  \\ %& 10   \\
\hline
\end{tabular}
\\
\tablefoottext{a}{The second number is the number of galaxies with masses above the completeness limit, with $10\leq \rm{log(M/M_{\odot})}\leq 11$.}\\
\end{table*}

%%%%%%%%%%%%%%%%%%%%%%%%%%%%%%%%%%%%%%%%%%%%%%%%%%%%%%%%%%%%%%%%%%%%%%%%%%

%
Members of each cluster were identified from the redshift versus projected cluster-centric radius plot as lying within
the “trumpet”-shaped caustic profile expected for galaxies infalling and subsequently orbiting within a massive virialized structure \citep[see details in][]{haines13,haines15}. 
A field sample was carefully determined to be  in narrow redshift slices either side of the clusters.  Because the properties of galaxies in the infall regions of the clusters 
are known to be systematically different from the field even at $3-4 R_{200}$ \citep[e.g.,][]{wetzel12}, we excluded from the field sample all galaxies within 4000\,km/s of the cluster \citep[see more details in][]{haines13}.

The selection of seven  out of 30 
%LoCuSS 
clusters was performed based on following criteria.
First, we excluded six clusters with $z>0.26$, 
with the [NII] EL  being redshifted  out of the observed spectrum range. We did not use four additional clusters without SDSS photometry. Additional thirteen clusters were excluded because of their much lower number ($<200$) of total spectroscopic members compared to the 
typical about $300-400$ (or even $>400$) spectroscopic members per cluster of this study (Tab.\,\ref{tab:7Clusters}). A small number of total spectroscopic members corresponds to a significant reduction of the number, in the mass-complete sample,  of cluster members with O/H measurements ($OHgals$ sample, see last column of Tab.\,\ref{tab:7Clusters}).
To avoid any bias due to such a small number of $OHgals$ objects per cluster, we excluded these thirteen clusters with $<200$ spectroscopic members from the present study.

\subsection{Star forming galaxies and type-2 AGNs}
\label{sec:type2AGNs}

\citet{cidfern11} introduced the WHAN diagram, $EW_{H\alpha}$ (equivalent width of H$\alpha$) vs. [NII]/H$\alpha$,  and used it to provide a comprehensive emission-line classification of SDSS galaxies. This diagram is useful to distinguish Type-2 AGNs from \emph{star-forming} galaxies with \emph{weak} [OIII] and H$\beta$ emission lines.
We used the WHAN diagram and followed the log([NII]/H$\alpha)$ criterium of \citet{gordon18}, who used the WHAN diagram to identify Type-2 AGNs for the GAMA survey.
Specifically, for our mass-complete cluster galaxy sample, we classified as $SFgals$ (star-forming galaxies) 410
star forming galaxies with a restframe $EW(H\alpha) >3$ \AA, a $S/N>5$ in  H$\alpha$ EL flux and log([NII]/H$\alpha)\leq -0.3$ , and as $SeyfLiners$ (Type-2 AGNs) 215
objects with a restframe $EW(H\alpha) >3$ \AA,  a $S/N>5$ in  H$\alpha$ EL flux and log([NII]/H$\alpha)>-0.3$ (see Fig.\,\ref{fig:WHAN}). The remaining 1340
cluster galaxies belong to the mass-complete $nonSFgals$ (non star-forming galaxies) sample.

The selection of the $SFgals$ sample is done using the WHAN diagram to ensure a complete sample of star-forming galaxies for comparison of the fraction of star-forming galaxies with theoretical models and simulations, including galaxies with stronger H$\alpha$, but weak [OIII] or H$\beta$.
On the other hand, we use the BPT diagram to select a sample $OHgals$ of galaxies with high enough S/N in [OIII] and H$\beta$ to derive metallicities. This is done as follows.
We first required a restframe $EW(H\alpha) >3$ \AA, a $S/N>5$ in  H$\alpha$ EL flux, a $S/N>2$ in each H$\beta$, [OIII] and [NII] EL fluxes, and a  restframe $EW(H\beta) >2$ \AA.  
From this sample we then excluded composite/AGN galaxies following the BPT diagram shown in Fig.\,\ref{fig:BPT}.
Taking error bars of the flux measurements into account, and to mimimize the number of excluded composite galaxies above but close to the demarcation line which are likely to have their H$\alpha$ line mostly produced by star-formation, 
we also include in the  $OHgals$ sample a few objects above and up to a maximum of 0.3\,dex to the right of the \citet{kaufm03} curve.
We remain with a sample of 259 galaxies in the mass-complete $OHgals$ cluster sample, and 169 galaxies in the mass-complete $OHgals$ field sample.
The numbers of $OHgals$ and composite/AGN galaxies in the different mass bins for the field and cluster samples are given in Tab.\,\ref{tab:subsamples}.

%###########################################################################################
\begin{table*}
\caption{Number of objects in different subsamples of cluster and field galaxies. See Sect.\,\ref{sec:type2AGNs} for details of the selection of the subsamples.}
\label{tab:subsamples}
\begin{tabular}{ccccccc}
\hline\hline      
Cluster sample &&&&&\\
&Parent sample
%\tablefootmark{a}
 &   $SFgals$   &   $nonSFgals$ & $SeyfLiners$ & $CompAGNs$  & $OHgals$ \\
\hline      
Spectroscopic redshifts                &2726 &   661  & 1792  & 273 & 87 & 421 \\
$\rm{10\leq log(M/M_{\odot})}< 10.5$    &943  &   269  & 599   & 75 & 29 & 174 \\
$\rm{10.5\leq log(M/M_{\odot})}\leq 11$ &1022 &   141  & 741   & 140 & 39 & 85 \\
$\rm{10\leq log(M/M_{\odot})}\leq 11$   &1965 &   410  & 1340  & 215 & 68 & 259 \\
\hline      
\hline
Field sample &&&&&\\
&Parent sample &   $SFgals$   &   $nonSFgals$ & $SeyfLiners$ & $CompAGNs$  & $OHgals$ \\
\hline      
Spectroscopic redshifts               &1068 &  443   & 452   & 173 & 66 & 284 \\
$\rm{10\leq log(M/M_{\odot})}< 10.5$    &307 &  146   & 106   & 55 & 24 & 125 \\
$\rm{10.5\leq log(M/M_{\odot})}\leq 11$ &398 &  81   & 232   & 85  & 29 & 44 \\
$\rm{10\leq log(M/M_{\odot})}\leq 11$   &705 & 227   & 338   & 140 & 53 & 169\\
\hline      
\hline
\end{tabular}
\\
\end{table*}

%############################################################

%##############################################################################

\subsection{Oxygen abundances, SFRs and stellar masses}
\label{Oxabund}

%########################################################################

\begin{figure*}[ht]
\includegraphics[width=10cm,angle=270,clip=true]{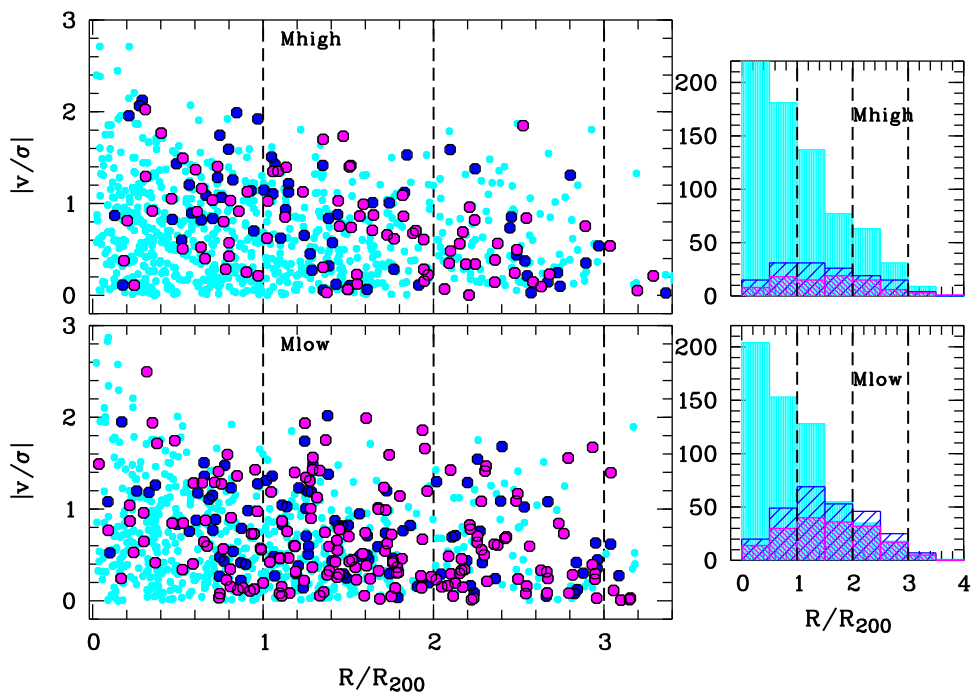}
\caption
{
\label{fig:PhaseSpLoCuSS} 
\footnotesize 
Phase-space diagram showing \emph{cluster} $SFgals$ (blue), $nonSFgals$ (cyan) and $OHgals$ (magenta) 
objects with $10 \leq \rm{log(M/M_{\odot})} \leq 11$. 
The histograms on the right for the different subsamples
as a function of clustercentric radius
are used to compute the observed fraction of SF galaxies and constrain quenching timescales by comparing with the Millennium simulation (cf. Fig.\,\ref{fig:fSF}).
The star-forming galaxies at $R<R_{200}$ (magenta and blue symbols) have in general higher line-of-sight velocities than non-SF galaxies, indicative of them to be an infalling population, being accelerated to high speeds as they pass through the cluster core.
}
\end{figure*}

%###########################################################################################
%########################################################################

\begin{figure*}[ht]
\includegraphics[width=12cm,angle=270,clip=true]{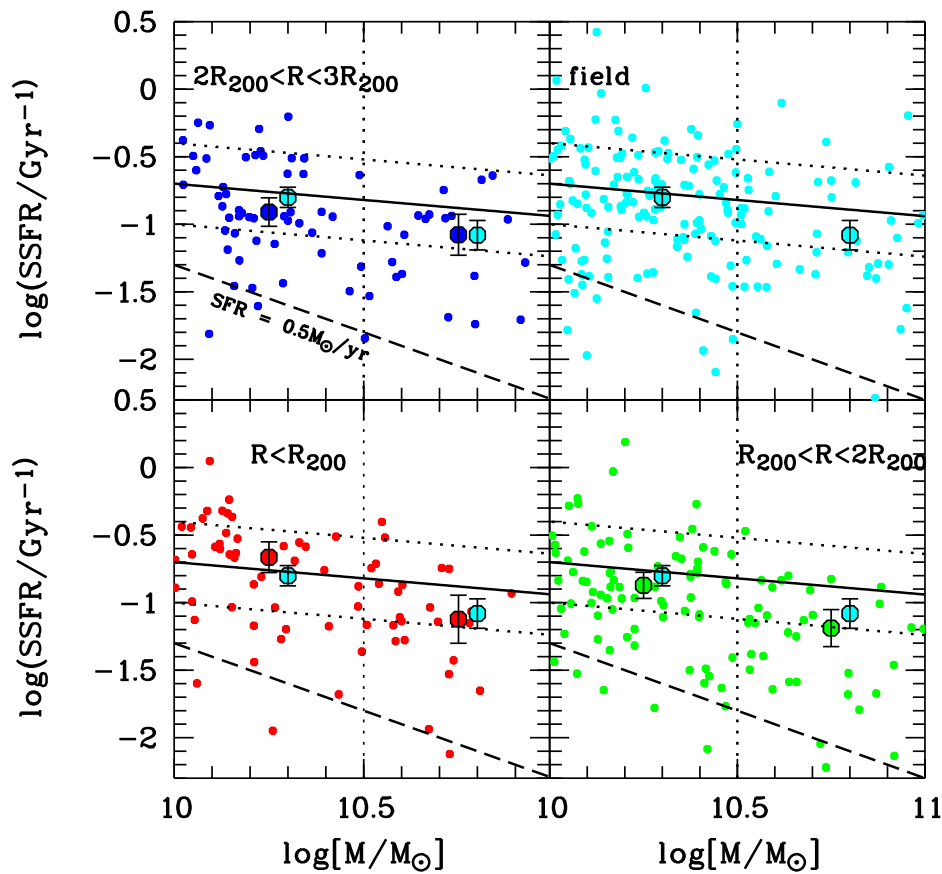}
\caption
{
\label{fig:MassSSFR} 
\footnotesize 
The mass$-$SSFR relation for 259 cluster and 169 field galaxies in the $OHgals$ mass-complete sample for three regions at different distances from the cluster center (red, green, blue) and field galaxies (cyan). The oblique solid line shows the main sequence  (MS) at $z \sim 0.2$ and its dispersion (indicated by the dotted lines) using Eq.\,1 in \citet{peng10}. The dashed oblique line in each panel shows 
that galaxies in our $OHgals$ sample have typically SFRs$>0.5M_{\odot}/yr$.
The median SSFRs in the Mhigh and in the Mlow  mass bins are shown as big filled circles, with the field median values (cyan symbols) shown in all panels (slightly displaced in mass compared to cluster galaxies for visibility of the median symbols).
In a given mass bin the median SSFR values of field and cluster galaxies are consistent.
%The mass-SSFR relation of star-forming galaxies in our $OHgals$ sample is rather independent of wheteher the galaxies are field or cluster galaxies, and also independent of cluster-centric radius.
% are hardly affected by environment.
}
\end{figure*}

%########################################################################

~~~After reduction of the Hectospec/MMT data using the SPECROAD pipeline \citep{mink07}, the EL fluxes of \OII, \Hb, \OIIIa, \Ha, and \NII\, for cluster member  and  field galaxies 
were measured (including stellar absorption correction for H$\alpha$ and H$\beta$) as described in detail in M.~J.~Pereira et al. \citep[in preparation, see also][]{pereira12}.
As in MKZ16, the Eq.\,4 in \citet{kewley13a}, $12+\rm{log}(O/H)=8.97-0.32 \times \rm{log}(([OIII]/\Hb)/([NII]/\Ha))$, which corresponds to the \citet{petpag04} O3N2 calibration for the \citet{kewdop02} metallicity scale, 
 was used to consistently compute metallicities for the cluster and field galaxies, and for the comparison SDSS sample. The SDSS sample was selected and matched as described in MKZ16 to ensure a consistent comparison of the physical properties of the SDSS and $OHgals$ samples.

%##############################################################################

~~~One of the most reliable and well calibrated SFR indicators is the
\Ha\, EL \citep{ken98}.
The conversion of \Ha\, luminosity to SFR requires several steps.
To correct for fiber losses for cluster galaxies 
we convolved each Hectospec spectrum  with the SDSS r-band filter and then compared this magnitude with the SDSS r-band magnitude of the respective galaxy. 
The difference between the two magnitudes gave the aperture
correction factor for each spectrum. 
This correction 
assumes that the  \Ha\, line flux and the r-band continuum suffer equal slit 
losses.
Thus revised  \Ha\, line luminosities $\rm{L}(\rm{H}\alpha)$ were corrected for extinction based on the Balmer decrement and then transformed into SFRs by applying  the \citet{ken98} conversion: $\rm{SFR} (M_{\odot}\rm{yr}^{-1}) = 7.9 \times 10^{-42}
\rm{L}(\rm{H}\alpha)\rm{ergs/s}$.
%

%######################################################################

~~~Stellar masses were computed from the available photometry using the g$-$i colour-dependent K-band mass-to-light relation  $log(M/L)_{K}=-1.321+0.754 \cdot (g-i)$ reported by \citet{zibet09}.
We converted the masses from a \citet{chabrier03} IMF  to a \citet{salp55} IMF (as used in this paper) by multiplying the \citet{chabrier03} IMF masses by 1.7. This factor of 1.7 was found by \citet{pozzetti07} to be a systematic median offset, with a very small dispersion, in the masses derived with the two different IMFs, being 
rather constant for a wide range of star formation histories.
%

%##############################################################################
%######################################################################

\section{Results: Environmental effects on the metallicities and SFRs of cluster galaxies}

%######################################################################
%%%%%%%%%%%%%%%%%%%%%%%%%%%%%%%%%%%%%%%%%%%%%%%%%%%%%%%%%%%%%%%%%%%%
\label{envOHs}

~~~The phase-space diagram (projected cluster-centric
radius vs. line-of-sight velocity relative to the cluster redshift) provides valuable information on the accretion history of cluster galaxies \citep[e.g.,][]{haines15}.
Fig.\,\ref{fig:PhaseSpLoCuSS} shows the phase-space diagram for the seven clusters, with R being the projected radial distance from the cluster center, and the different symbols indicating different galaxy subsamples, $nonSFgals$ (cyan), $SFgals$ (blue) and $OHgals$ (magenta). Note that the $OHgals$ sample is a subset of the star-forming sample, $SFgals$.
The histograms on the right with the number of objects as a function of clustercentric radius in different subsamples are later used (in Sect.\,\ref{sec:observed_fSF}) to compute the fraction of star-forming galaxies  and to derive quenching timescales.

As shown in Fig.\,13 of \citet{haines15}, infalling galaxies which have recently passed within $R_{200}$ for the first time, but have yet to reach the pericenter, have high line-of-sight velocities, being accelerated as they fall deep into the gravitational potential well of the cluster core. The star-forming galaxies at $R<R_{200}$ (magenta and blue symbols) shown in the phase-space diagram (Fig.\,\ref{fig:PhaseSpLoCuSS}) have in general higher line-of-sight velocities than non-SF galaxies (cyan filled circles), indicative of them to be an infalling population.

For a meaningful investigation and  considering our 
sample of targets (e.g., only seven cluster galaxies in the $OHgals$ sample have $\rm{log(M/M_{\odot})} > 11$), we divide the galaxies in our mass-complete sample into two mass-bins:
i) galaxies with $Mlow$ 
($10 \leq \rm{log(M/M_{\odot})} < 10.5$), and ii) galaxies with 
$Mhigh$
($10.5  \leq \rm{log(M/M_{\odot})} \leq 11$). 
%

%######################################################################
%%%%%%%%%%%%%%%%%%%%%%%%%%%%%%%%%%%%%%%%%%%%%%%%%%%%%%%%%%%%%%%%%%%%

\subsection{The 
mass$-$SSFR relation in field and clusters at $z \sim 0.2$}
\label{SSFRMsel}

%########################################################################

\begin{figure*}[ht]
%MassOHMED_7ClustersEnvDistDeltaSEP18.ps
%MassOHMED_7ClustersEnvDistDeltaOCT18.ps
\includegraphics[width=13cm,angle=270,clip=true]{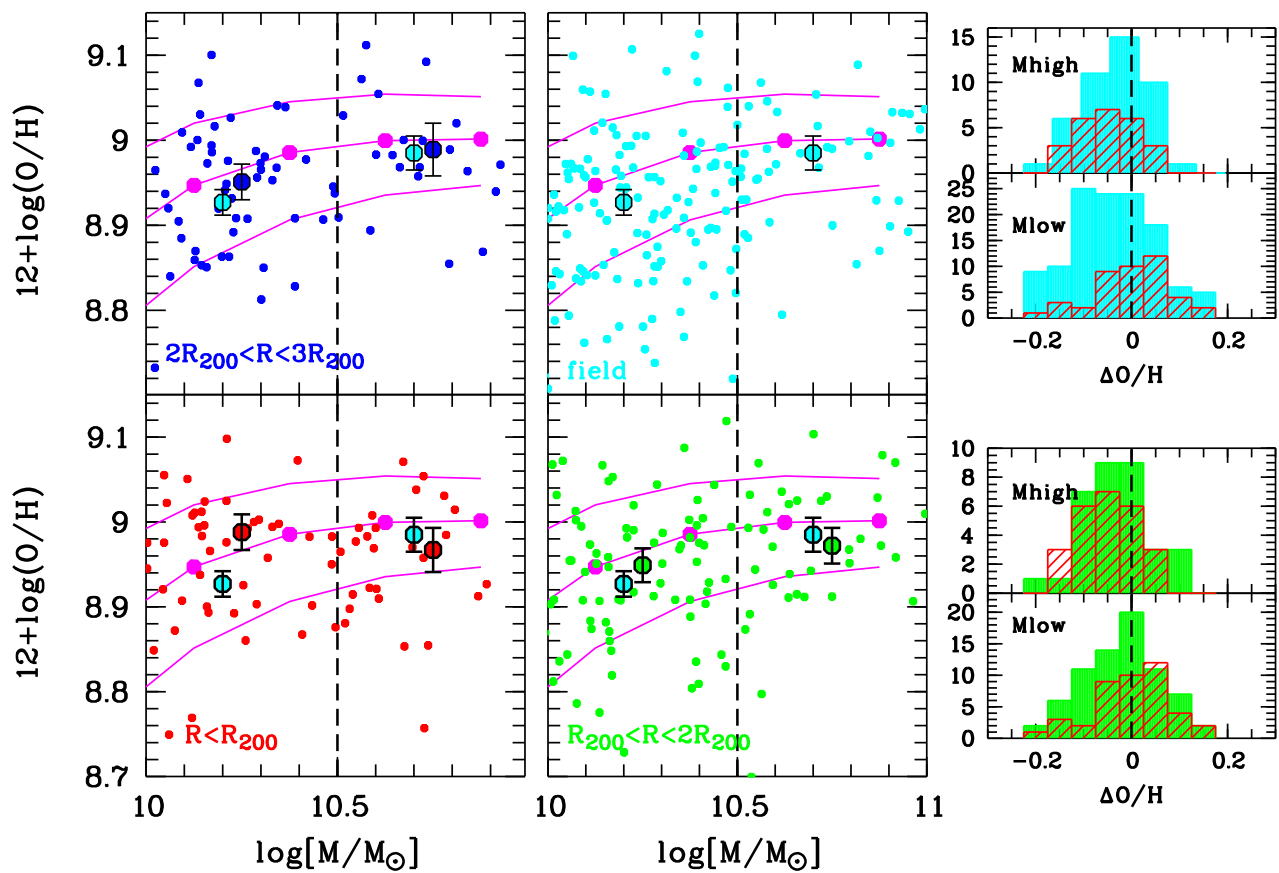}
\caption
{
\label{fig:MassOH} 
\footnotesize 
MZR and dependence on environment of the 259 cluster galaxies and 169 field galaxies of the mass-complete $OHgals$ sample.
%MZR and dependence on environment for the 421 cluster galaxies and 284 field galaxies of the \emph{entire} $OHgals$ sample.
For the matched SDSS sample, we indicate 16th, and 84th percentiles, and the medians (50th percentiles) of the distribution of O/H values in the respective mass bin as magenta lines and symbols. The median O/Hs in the mass-complete Mhigh and Mlow mass bins 
%(and in the not mass-complete $9.5 \leq \rm{log(M/M_{\odot})}<10$ mass bin) 
are shown as big filled circles, with the field median values (cyan symbols) shown in all panels (slightly displaced in mass compared to cluster galaxies for visibility of the median symbols).
The field median O/H value of Mhigh galaxies of the LoCuSS sample is similar to the SDSS median value at higher masses, while the Mlow field median  value is slighty lower than the Mlow SDSS median value. The field Mlow and Mhigh median metallicities values are 
 comparable to the respective (Mlow, Mhigh) cluster O/H medians for $R>R_{200}$. On the other hand, the median O/H value of Mlow galaxies in the more central part of the clusters ($R<R_{200}$) is higher than the median O/H field value, by about 0.06\,dex, with a $2.4\sigma$ significance.
The histograms on the right showing the offsets from the local SDSS MZR relation, $\Delta O/H$,  also clearly reveal enhanced metallicities for Mlow cluster galaxies inside $R_{200}$ (red histograms) compared to field galaxies and to  galaxies at $R>R_{200}$. 
This indicates that the cluster environment has the effect of increasing the gas metallicities of Mlow galaxies inside $R_{200}$ compared to their field counterparts of similar masses.
}
\end{figure*}
%########################################################################

~~~\citet{peng10} derived a formula of the evolution of the SSFR as a function of mass and time that we use to calculate the mean SSFR as a function of stellar mass at $z \sim 0.2$. For this, we assume a dependence of the SSFR on mass as observed for local SDSS galaxies and revised by \citet{renzpeng15}, SSFR$\propto$M$^{\beta}$ with $\beta=-0.24$ (Fig.\,\ref{fig:MassSSFR}, oblique solid black line).  
The local main sequence (MS) relation was found  to have a dispersion of a factor of 0.3\,dex 
about the mean relation \citep[e.g.,][]{salim07,elbaz07,peng10}, 
indicated by the dotted oblique lines
around the mean MS relation at $z \sim 0.2$ in Fig.\,\ref{fig:MassSSFR}.

~~~In a given mass bin (Mlow or Mhigh) the median SSFR values of field and cluster galaxies are consistent, given their error bars. 
Thus, the mass-SSFR relation of star-forming galaxies in our $OHgals$ sample is rather independent of wheteher the galaxies are field or cluster galaxies, and also independent of cluster-centric radius.
%environmental effects
%strongly affecting the SFRs of cluster \emph{star-forming} galaxies are not visible.
This is consistent with the MKZ16 results studying the MACSJ0416.1-2403 cluster and finding a similar distribution of \emph{star-forming} cluster and field galaxies in the SSFR-mass diagrams.

%######################################################################
\subsection{The enhanced metallicities of cluster galaxies inside $R_{200}$}
\label{sect:MZR}

 ~~~Fig.\,\ref{fig:MassOH} shows the MZR of  the $z \sim 0.2$ cluster and field galaxies.
All metallicities were computed with the same method, as described in Sect.\,\ref{Oxabund}.
The field median Mlow and Mhigh metallicities (cyan symbols) are 
similar 
%to the SDSS median metallicity and 
to the $R>R_{200}$ cluster median metallicity at a given mass (Mlow or Mhigh, respectively). On the other hand, the median metallicity of Mlow galaxies in the higher density parts of the clusters ($R<R_{200}$, red symbol) is 
higher than the median metallicity of Mlow \emph{field} galaxies, by about 0.06\,dex,  with a $2.4\sigma$ significance.

The distribution of the metallicities of $R<R_{200}$ Mlow galaxies (small red filled symbols) is clearly shifted to higher values compared to their field counterparts (small cyan symbols). This is also seen in the histograms of the O/H offsets from the local SDSS MZR relation $\Delta O/H$ (shown on the right in Fig.\,\ref{fig:MassOH}) which clearly depict enhanced metallicities of Mlow cluster galaxies  inside $R_{200}$ (red histogram) compared to field galaxies (cyan histogram) and to  galaxies at $R>R_{200}$ (green histogram). At higher masses Mhigh the metallicities of cluster and field galaxies broadly agree, with no sign of strong environmental effects on the O/H values. We discuss in the next section the reason for this behaviour at Mhigh, where the MZR saturates.
%

%Recently, \citet{quai18} showed that galaxies with lower [OIII]/H$\beta$ are the best candidates for galaxies in the act of quenching star formation. 
%
%The histograms in the right panels of Fig.\,\ref{fig:BPT} show that Mlow galaxies inside $R_{200}$ (red histogram), which generally have \emph{enhanced metallicities} compared to $R>R_{200}$ objects (Fig.\,\ref{fig:MassOH}), typically have \emph{lower [OIII]/H$\beta$ ratios}  compared to $R>R_{200}$ galaxies (green histogram) and field objects (cyan histogram in Fig.\,\ref{fig:BPT}). 
%This implies that the higher gas-phase metallicities we observe for Mlow galaxies inside $R_{200}$ in general also correspond to lower [OIII]/H$\beta$ ratios.
%
%Combined with the findings of \citet{quai18}, we can conclude that Mlow galaxies inside $R_{200}$ with increased gas metallicities are galaxies \emph{in the act of quenching star formation}.

%######################################################################

\section{Discussion: Strangulation and slow-then-rapid quenching}
\label{delayrapid}

%######################################################################
%%%%%%%%%%%%%%%%%%%%%%%%%%%%%%%%%%%%%%%%%%%%%%%%%%%%%%%%%%%%%%%%%%%%

%########################################################################

\subsection{Strangulation implied by the comparison to the Z(M$_{*}$,SFR) expectations with inflow of pristine gas}
\label{sect:MZR}

%########################################################################

\begin{figure*}[ht]
\includegraphics[width=10cm,angle=270,clip=true]{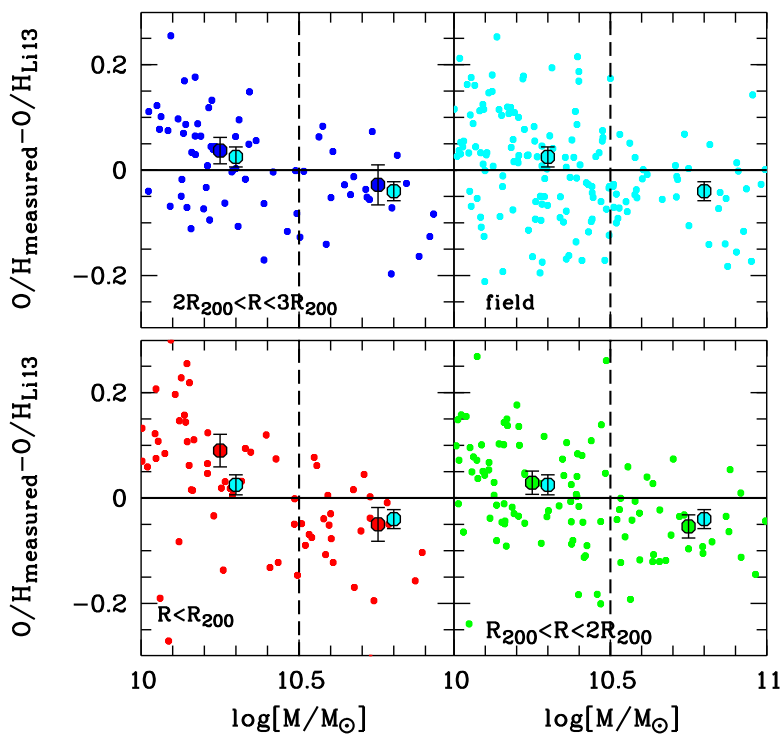}
\caption
{
\label{fig:DiffLi13} 
\footnotesize 
Difference between the measured O/Hs of $z \sim 0.2$ cluster galaxies and the expected O/Hs from the formulations of \citet{lilly13} for three regions at different distances from the cluster center (red, green, blue) and field galaxies (cyan). 
The median value of the differences in metallicities of  Mlow cluster galaxies inside $R_{200}$ (large red filled circle) indicates 
%with a 3$\sigma$ significance, 
that measured O/Hs are higher inside $R_{200}$ compared to the \citet{lilly13} predictions with inflow of pristine gas. On the other hand, the median O/Hs of $Mlow$ field galaxies (large Mlow cyan filed circle) and the median O/Hs of $Mlow$ galaxies at $R>R_{200}$ are quite consistent (1.3$\sigma$) with the predictions. 
This indicates that the environment has an influence on the gas metallicities with slow quenching (strangulation) likely initiated when galaxies pass $R_{200}$.
}
\end{figure*}

%########################################################################

~~~To further explore the reason for the enhancement of O/Hs for Mlow galaxies in the cluster environment we calculate the expected O/H values from the \citet{lilly13} model for each galaxy individually with their respective stellar mass M$_{*}$ and SFR.
\citet{lilly13} proposed a simple model of galaxy evolution in which the SFR is regulated by the mass of gas present in a galaxy, implying that Z depends on both  M$_{*}$ and current SFR. 
Their derived Z(M$_{*}$,SFR) relation has a particular analytic form:
the equilibrium metallicity  $Z_{eq}$ in the gas regulator model is given by Z$_{eq}=\rm{Z}_{0}+$f$_{star}$y, with  $Z_{0}$  the metallicity of the infalling gas, and  $y$ the yield. $f_{star}=SFR/\Phi$ is the fraction of the incoming baryons (inflow $\Phi$) that are converted into long-lived stars.
We use their derived Z(M$_{*}$,SFR) relation assuming \emph{primordial inflowing gas} ($Z_{0}=0$) in their regulator bathtub model.
We then compare the \emph{predicted} values with the \emph{measured} ones and show the difference  in Fig.\,\ref{fig:DiffLi13}.
Studies of blue supergiants \citep{bresolin16} and red supergiants \citep{davies17} showed that the strong-line calibration yielding the most accurate (absolute) metallicities is the O3N2 calibration which we use. Therefore, our measured O3N2 metallicities should be well suited to be compared,  in an absolute way, with the metallicities derived using the \citet{lilly13} model.

The median values depicted in Fig.\,\ref{fig:DiffLi13} show that Mlow cluster galaxies inside $R_{200}$ (red) have  higher metallicities
%, with a $3\sigma$ significance, 
compared to the \citet{lilly13} predictions with inflow of \emph{pristine} gas, while Mlow field galaxies and galaxies at $R>R_{200}$ are in general consistent with these predictions as shown by their median values.
This indicates a \emph{strangulation scenario} like discussed by MKZ16 (by studying the cluster MACSJ0416.1-2403).
If the SFR is hardly affected (as discussed in Sect.\,\ref{SSFRMsel}), a higher value for the metallicity $Z_{eq}$ in the above equation can be only obtained by decreasing or stopping the inflow of pristine gas $\Phi$ (in the denominator in the above equation).
The inflow of pristine gas onto the galaxy disk is likely halted when Mlow cluster galaxies are accreted inside $R_{200}$, and their gas-phase metallicities thus increase, because  their interstellar medium is no longer diluted by the inflow of pristine gas.

The MZR saturates when the mass of oxygen produced by massive stars (which goes into the interstellar medium) equals the mass of oxygen locked up by lower-mass longer-living stars. For $Mhigh$ in the LoCuSS sample, the MZR saturates and becomes flat for higher masses, as seen in Fig.\,\ref{fig:MassOH} for all environments \citep[see also discussion in][]{zahid14}. Therefore, strangulation does not have any visible (measurable) effect on 
metallicities of $Mhigh$ objects. 

The
quenching (strangulation) is likely initiated when galaxies pass $R_{200}$, but this first only affects the gas metallicities, not much the SSFRs of cluster objects (see Fig.\,\ref{fig:MassSSFR}). 
The galaxies then slowly use up their existing 
%molecular 
gas reservoir through star formation.
%, possibly over \textbf{a period of about $2-3$\,Gyrs}, according to  the constant molecular gas consumption timescales observed for the disks of nearby spiral galaxies spanning a wide range of properties \citep{bigiel11}. 
However, because galaxies inside $R_{200}$ can travel to denser inner regions of the cluster in about 1\,Gyr, it is probable that ram-pressure stripping processes will be more pronounced with time, contributing to a \emph{rapid complete quenching} of star-formation at a later time, as we discuss in Sect.\,\ref{sec:slowrapid}.
%

%##############################################################################

\subsection{$f_{SF}$ predictions from simulations for different time delays between accretion and quenching}
\label{sec:modelSF}

%#########################################################
\begin{figure*}[ht]
\includegraphics[width=9cm,angle=270,clip=true]{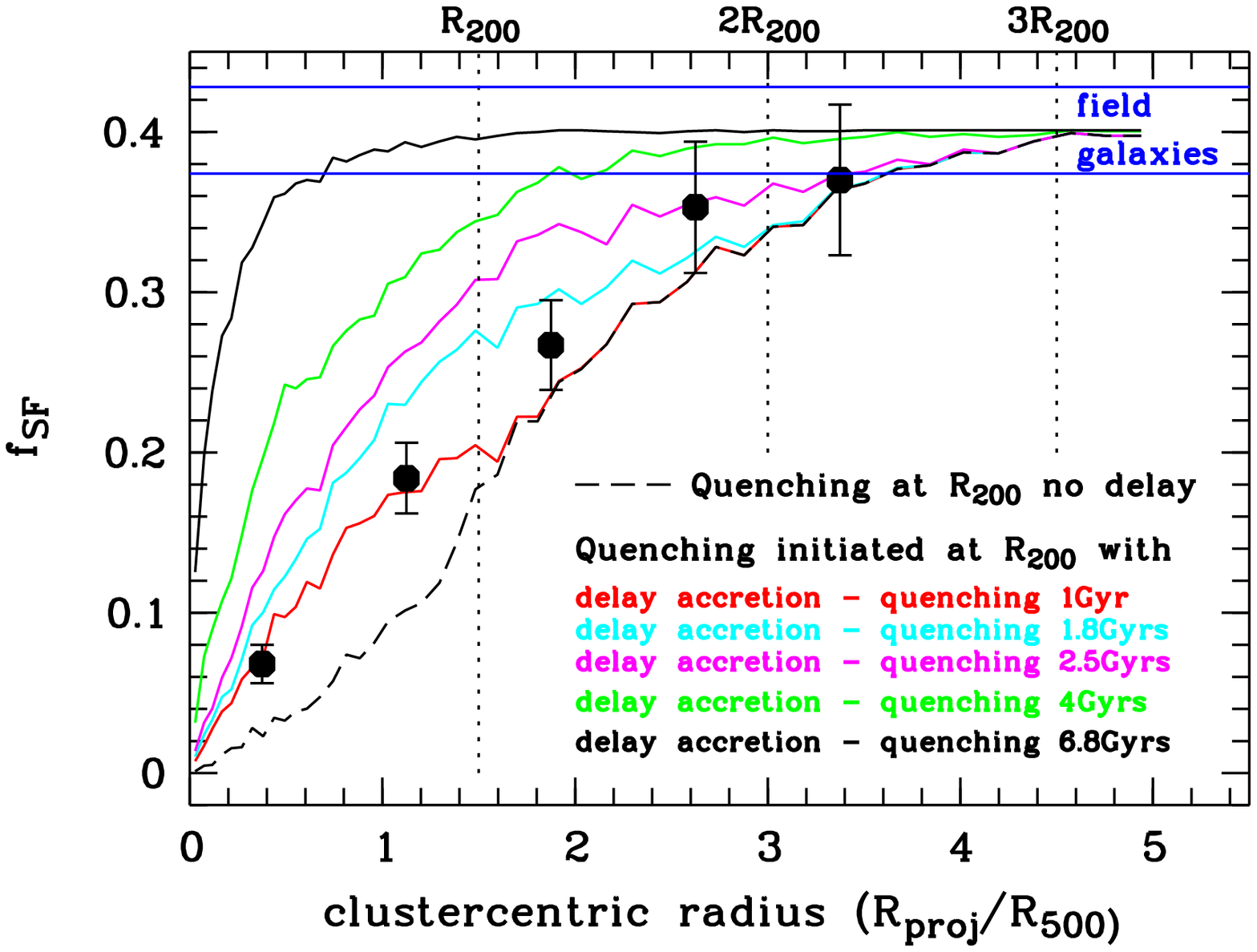}
\caption
{
\label{fig:fSF} 
\footnotesize 
Comparison of the observed fraction of SF galaxies $f_{SF}$  as a function of clustercentric radius (based on the histograms shown in Fig.\,\ref{fig:PhaseSpLoCuSS})
with predictions from the Millennium cosmological simulation. The black filled circles show the observed fraction of SF cluster galaxies in the LoCuSS mass-complete sample.
Each galaxy is weighted by the inverse probability of it having being observed spectroscopically,
following the approach of \citet{norberg02}.
 The black \emph{dashed} line shows the predicted SF-radius relation using 75 massive clusters in the Millennium simulation, assuming that star formation is immediately quenched upon being accreted by the cluster, i.e., passing the physical $R_{200}$ 
for the first time. 
The additional (colored) solid curves show the effects of delaying the \emph{complete} quenching by a time $\Delta t$ after the galaxy is accreted into the cluster  (i.e., passing $R_{200}$ for the first time), corresponding to the ``delayed-then-rapid'' quenching scenario of \citet{wetzel13} or to a "slow-then-rapid'' model as discussed in Sect.\,\ref{sec:slowrapid}. The comparison of the observed fractions with the model curves derived from the Millennium simulation implies $\Delta t \sim1-2$\,Gyrs.
}
\end{figure*}

%#########################################################

~~~To constrain the timescales on which star formation is quenched as galaxies fall into clusters, e measure how the observed fraction of star-forming galaxies increases with cluster-centric radius (the SF-density relation, see histograms in Fig.\,\ref{fig:PhaseSpLoCuSS}), and compare it with predictions from cosmological simulations, adopting the same approach as \citet{haines15}.
They examined  the spatial
distributions and orbits of galaxies in the vicinity of the 75 most massive ($3-21 \times 10^{14}M_{\odot}$) clusters from the Millennium simulation \citep{springel05}, 
%same cosmology 0.25, 0.75 as we use in this paper !
a cosmological dark matter simulation covering a (500$h^{-1}$Mpc)$^{3}$ volume. 

~~~Twenty$\times 20 \times 140h^{3}$Mpc$^{3}$ volumes centered on each cluster were extracted and extended in the z-direction so that, for a distant observer viewing along this axis, all galaxies with line-of-sight velocities within 5000\,kms$^{-1}$ of the cluster redshift are included, enabling projection effects to be fully accounted and quantified.
Using a database with information on simulated galaxy properties and dark matter halos information \citep[see more details in][]{haines15} one can follow the orbit of each galaxy with respect to the cluster from formation to the present day, enabling to determine its epoch of accretion into the cluster defined here as the redshift at which the galaxy passes within $R_{200}$ for the first time. The backsplash galaxies discussed below are still hosted by dark-matter halos, so their positions and orbits are well defined, simply by tracking their host dark-matter halo. An important point is that we take into account projection effects, since we are looking through the cluster, not at it. Thus we include all the galaxies along the line-of-sight, but which would be identified as spectroscopic cluster members since they are within the wider overdense region.

The stacked radial population gradients for 75 massive 
% ($3-21 \times 10^{14}M_{\odot}$) 
clusters from the Millennium simulation are reproduced, as they would appear if observed at 
$z \sim 0.2$, to best match the redshifts of the LoCuSS sample.
% \citep[see details in][]{haines15}.
% 
The model galaxy positions and velocities relative to the cluster halo are measured as they stood at $z \sim0.2$, and the clusters are stacked using their $R_{200}$ values measured at $z \sim 0.2$.
The form of the black dashed curve in Fig.\,\ref{fig:fSF} is mostly due to the large population of galaxies in the wider over-density around the cluster that are still on their first infall, but still outside $R_{200}$, and which become increasingly important in terms of numbers for increasing clustercentric radius, combined with the backsplash galaxies whose orbits back out beyond $R_{200}$.
In this simple toy model we assume that quenching is initiated when  infalling field galaxies pass physical $R_{200}$ of the cluster for the first time, based on our observational results of enhanced metallicities at $R<R_{200}$ and initiated quenching at $R_{200}$ and also in agreement with \citet{wetzel12} who concluded that infalling massive galaxies experience no significant quenching of star-formation
%environmental effects 
until they arrive within $R_{200}$ of the more massive host halo.
It should be noted that the observational (tentative) evidence for quenching initiated when galaxies pass $R_{200}$ for the first time is a new result from our analysis, and was not known in the work of \citet{haines15} who also considered models in which star formation starts to be quenched when galaxies pass $R_{500}$ or $2R_{200}$ for the first time (their Figs.\,14 and 15).

This onset of quenching could result either in the star formation of infalling galaxies being \emph{instantaneously completely quenched} 
%\textbf{(moving them from the $SFgals$ to the $nonSFgals$ sample)} 
at the moment they pass physical  $R_{200}$, or \emph{completeley quenched} after a \emph{time delay} meanwhile \emph{slow} (but not \emph{complete})  quenching can already happen. ``Complete'' quenching means for our observational sample that galaxies leave the $SFgals$ and enter the $nonSFgals$ sample.
To consistently compare the observed relation with the models, the fraction of SF galaxies $f_{SF}$ vs. clustercentric radius relation  shown in Fig.\,\ref{fig:fSF} assumes that the population of infalling galaxies has the same $f_{SF} \sim 40$\% as our observed field galaxy sample (blue solid lines in Fig.\,\ref{fig:fSF}), i.e., 40\% of field infalling galaxies are in the $SFgals$ sample and 60\% in the $nonSFgals$ sample. 

We note that this $f_{SF}$ vs. clustercentric radius diagram cannot distinguish
between a delayed-then-rapid quenching and a slow-then-rapid quenching scenario.
It just indicates how long galaxies remain classified as star-forming ($SFgals$) before they are quenched becoming $nonSFgals$. Their star-formation may be hardly affected during this ``delay'' time or slowly declining, while they remain classified as $SFgals$.

For $R_{200}<R<2R_{200}$ the ``instantaneous quenching'' model $f_{SF}$ curve (dashed line in Fig.\,\ref{fig:fSF}) does not reach the field  $f_{SF}$ value of about 40\% (as indicated by the area between the blue solid lines).
This is because of the contribution of \emph{quenched} ``back-splash'' galaxies that had been previously accreted into the cluster, reaching well within $R_{200}$ on first pericentric passage at earlier times and  quenching star formation, but now have bounced back out beyond $R_{200}$ \citep[e.g.][]{gill2005,ludlow09}. 
We note that \emph{quenched} ``back-splash'' galaxies appear as $nonSFgals$ in our LoCuSS sample.
The solid model curves show the effects of delaying the moment at which complete quenching occurs, by terminating star formation only in those galaxies accreted into the cluster more than $\Delta t$ Gyrs prior to observations, with $\Delta t$ corresponding to the delay time in the ``delayed-then-rapid'' quenching scenario of \citet{wetzel13}.
The red curve (1\,Gyr delay between accretion
and complete quenching) diverges from the dashed curve (instanteneous rapid quenching when accreted at $R_{200}$)
only for  $R<R_{200}$, because the most recently accreted galaxies (less than 1\,Gyr ago) have not had sufficient time to pass through the clusters and go back beyond $R_{200}$. 
%

%%%%%%%%%%%%%%%%%%%%%%%%%%%%%%%%%%%%%%%%%%%%%%%%%%%%%%%%%%%%%%%%%%%%
\subsection{Observed $f_{SF}$ and implied quenching timescales}
\label{sec:observed_fSF}

~~~The observed fraction  $f_{SF}$ of star forming galaxies in our cluster and field samples was computed based on the classification as $SFgals$ using spectroscopic \Ha-measurements (as described in Sect.\,\ref{sec:type2AGNs}), which measure instantaneous SFRs (timescale $< 10$Myr), different from the LoCuSS classification based on 24 $\mu m$ SFRs  by \citet{haines13,haines15}.  
Thus, our SFgals sample is not a subsample of the sample used by \citet{haines13,haines15}.
The SFRs from 24 $\mu m$  have significant contributions from slightly older populations \citep[as shown by, e.g.,][]{salim09}, with timescales over which the star formation is measured being effectively on the order of 
%(a few) 
100\,Myrs.
Thus, using $f_{SF}$ based on \Ha-measurements we can measure quenching timescales with a complementary SFR indicator compared to 24 $\mu m$, which was used by  \citet{haines15} to derive quenching timescales. Using \Ha\, which measures SFRs on shorter timescales than 24 $\mu m$, we are able to detect the quenching of star-formation after a shorter time (i.e., more accurate).

The likelihood that a given galaxy was targeted for spectroscopy depends strongly on both its location with respect to the cluster center as well as its photometric properties (e.g., K-band magnitude, $J-K$ color) 
as detailed in \citet{haines13}. To account for this, each galaxy shown in the phase-space diagram and histograms in Fig.\,\ref{fig:PhaseSpLoCuSS} is additionally weighted by
the inverse probability of it having been observed spectroscopically,
following the approach of \citet{norberg02}.
The resulting $f_{SF}$ are shown as black filled circles in Fig.\,\ref{fig:fSF}.
Error bars for the observed fraction of SF galaxies $f_{SF}$ are computed using Poisson statistics $\Delta f_{SF} = f_{SF} \cdot \sqrt{1/N_{SF}+1/N_{SFnonSF}-2/N_{SF}\cdot N_{SFnonSF}^{-1/2}}$\,, where $N_{SF}$ is the number of SF galaxies 
and $N_{SFnonSF}$ is the total number of SF and non-SF galaxies.
The comparison of the observed $f_{SF}$ (black symbols) with the model curves derived from the Millennium simulation
implies quenching timescales of 1$-$2\,Gyrs, roughly consistent with a location between the red and cyan model curves, given the error bars. Much longer quenching time-delays ($\Delta t > 3$\,Gyrs, green curve) are clearly excluded, because they would leave too many star-forming galaxies at $R<2R_{500}$, which is not observed. These quenching time-delays of 1$-$2\,Gyrs are consistent with the delay times for cluster galaxies with similar stellar masses  ($10\leq \rm{log(M/M_{\odot})}\leq 11$) found  by \citet{balogh16}, as shown in their Fig.\,7, and also with the quenching timescales of $1-2$\,Gyrs inferred by \citet{haines13, haines15} for galaxies accreted inside $R_{200}$.

%%%%%%%%%%%%%%%%%%%%%%%%%%%%%%%%%%%%%%%%%%%%%%%%%%%%%

\subsection{Slow - traced by enhanced metallicities - then rapid quenching}
\label{sec:slowrapid}
%
%%%%%%%%%%%%%%%%%%%%%%%%%%%%%%%5

~~~ \citet{haines13} found that massive star-forming galaxies are slowly quenched upon accretion into massive clusters, consistent with a gradual shut
down of star formation in infalling spiral galaxies as they interact with the ICM via ram-pressure stripping or strangulation mechanisms.
The SFRs of these infalling galaxies were found to decline 
exponentially on quenching timescales in the range $0.7-2.0$\,Gyr.

~~~ On the other hand, \citet{wetzel13} concluded that the SFRs of satellite 
cluster galaxies evolve, at least on average, via a ``delayed-then-rapid'' quenching scenario. The SFRs of lower mass satellites remain unaffected for a delay of a few Gyrs  after first infall, and this delay timescale was claimed by  \citet{wetzel13} to be shorter for more massive satellites. This is consistent with $\Delta t \sim1-2$\,Gyrs we found for $10\leq \rm{log(M/M_{\odot})}\leq 11$ cluster galaxies (Fig.\,\ref{fig:fSF}).
After the delay-phase, the SFRs of accreted satellite galaxies is expected to decline rapidly in the \citet{wetzel13} model.

This \emph{delayed}-then-rapid quenching scenario is compatible with our findings if changed to a \emph{slow}-the-rapid quenching scenario, with a significant difference in the interpretation compared to \citet{wetzel13}. 
Slow quenching (strangulation) is initiated when infalling galaxies first pass $R_{200}$ and traced by an enhacement of O/Hs of Mlow galaxies (due to the stopping of gas inflow) for a (delay) time $\Delta t \sim1-2$\,Gyrs.

\citet{maier06} constructed a large grid of P\'egase2 models to explore which region of the parameter space could reproduce the constraints imposed by the local
metallicity-luminosity relation and by the metallicities and luminosities of galaxies at higher redshifts. 
They explored models in which the gas inflow on the model galaxy stops and the galaxy subsequently forms stars in a 
``closed-box''-like scenario.
The tracks of these closed-box models show that, after the gas inflow stops,
galaxies increase their metallicities by $\sim 0.2$\,dex in   $t_{\rm{enrich}}\sim 1$\,Gyr.
Thus, strangulated infalling galaxies can increase their metallicities  by $\sim 0.2$\,dex while they move towards the central region of the cluster.
This enrichment time  is smaller than the typical crossing time $t_{\rm{crossing}}=R_{200}/\sigma$ of our studied clusters of about 2\,Gyrs.
During strangulation the inflow of new gas is cut out, but the cold interstellar medium disk is not directly perturbed. In this case the star formation can continue, using the gas available in the disk (resulting in higher gas-phase metallicities) until it is completely used up, unless an additional effect quickly completely quenches star-formation before.

%%%%%%%%%%%%%%%%%%%%%%%%%%%%%%%%%%%%%%%%%%%%%%%%%%%%%%%%%%%%%%%%%%%

Our observational results revealing enhanced metallicities inside $R_{200}$ and the quenching timescales that we derive by comparing the observed $f_{SF}$ with theoretical models and simulations in Fig.\,\ref{fig:fSF} can constrain and identify a likely physical quenching mechanism. This physical mechanism has to be active around $R_{200}$ to stop the infall of pristine gas on the galaxies. The galaxies then remain in the $SFgals$ sample (see Fig.\,\ref{fig:MassSSFR}) in a ``slow-quenching phase'', while they travel to denser inner regions of the cluster. They slowly use up their existing molecular and HI gas through star formation over a period of about $2-3$\,Gyrs, according to  the constant molecular gas consumption timescales observed for the disks of nearby spiral galaxies spanning a wide range of properties \citep{bigiel11} and based on the gas consumption timescales of star-forming galaxies  derived from their measured HI and molecular gas masses in our studied LoCuSS A963 cluster \citep{jaffe15,cybulski16}.
% in about 1\,Gyr. 
However, our derived quenching timescales of $1-2$\,Gyrs indicate that an additional effect has to quickly completely quench star-formation before the existing gas  is used up in $2-3$\,Gyrs, to move galaxies from the $SFgals$ to the $nonSFgals$ sample producing the observed $f_{SF}$ vs. clustercentric radius distribution shown in Fig.\,\ref{fig:fSF}. The physical mechanism producing this ``fast quenching'' has to be active when galaxies arrive in denser inner regions of the clusters, where the density of the ICM is larger. This dependence of the quenching mechanism on the density of the ICM (which increases with decreasing clustercentric radius of the clusters) makes ram-pressure stripping a promising candidate for the ``slow-then-rapid quenching" scenario, and we quantitatively explore this in the following.

%%%%%%%%%%%%%%%%%%%%%%%%%%%%%%%%%%%%%%%%%%%%%%%%%%%%%%%%%%%%%%%%%%%

%
%After $\Delta t \sim1-2$\,Gyrs 
%meanwhile the galaxies travel to denser inner regions of the cluster, also the SFRs of galaxies have to be considerably affected and 
%to decline quickly being completely quenched, to reproduce the results seen in Fig.\,\ref{fig:fSF}. 
%
Gas can be removed from infalling galaxies if the ram pressure exceeds the restoring force per unit area (gravitational restoring pressure) exerted by the galaxy, as first derived by \citet{gunngott72}. Using high-resolution cosmological hydrodynamic simulations, \citet{bahe13} computed the density of the ICM and derived the ram-pressure $P_{ram}$ in clusters. For massive clusters ($\sim 10^{15}M_{\odot}$) like the clusters we are studying, they estimated a range $P_{ram} \sim 3 \times 10^{-14} N/m^{2}- 10^{-12} N/m^{2}$ 
near $R_{200}$ for 90\% of the simulated galaxies \citep[Fig.\,9 in][]{bahe13}.
On the other hand, the typical restoring pressure derived by \citet{bahe13} for massive galaxies ($10 \leq \rm{log(M/M_{\odot})} \leq 11$) is  $\sim 3 \times 10^{-12} N/m^{2} - 10^{-11} N/m^{2}$ on cold gas and  $\sim 3 \times 10^{-14} N/m^{2}$ on hot gas at a radial distance from the galaxy center enclosing 0.1$M_{*}$ in cold gas. The reason for this restoring pressure difference is that cold gas is denser and sits much closer to the galactic center.

Comparing the ram-pressure $P_{ram}$ at $R_{200}$ with the restoring pressure, it is clear that \emph{the ram-pressure is too low at $R_{200}$ to strip cold gas, but becomes just sufficient ( $>3 \times 10^{-14} N/m^{2}$) for most infalling galaxies to strip the more extended, less dense and less tightly bound hot gas}.
Thus, while not impacting star formation directly, the removal of hot gas at $R_{200}$ ends the replenishment (through cooling) of cold gas resulting in a delayed decrease in star formation as the remaining cold gas is consumed (strangulation). At a later time, meanwhile the galaxies travel to denser inner regions of the cluster, $P_{ram}$ increases to values $P_{ram} > 3 \times 10^{-12} N/m^{2}$ comparable to the restoring pressure of cold gas, sufficient to strip \emph{cold} gas in massive cluster galaxies, producing a ``rapid'' phase of (complete) quenching of star formation. 

This interpretation is also consistent with the MKZ16 results about strangulation traced by enhanced metallicities initiated at $ \sim 2R_{200}$ (Fig.\,7 in MKZ16) for lower mass, $9 \leq \rm{log(M/M_{\odot})} \leq 10$, galaxies in the massive $10^{15}M_{\odot}$ MACSJ0416.1-2403 cluster. At $2R_{200}$ the ram-pressure calculated by \citet{bahe13}, $P_{ram} > 3 \times 10^{-15} N/m^{2}$,
becomes just comparable with the restoring pressure of the hot gas in the range $10^{-14} N/m^{2} - 10^{-15} N/m^{2}$ of lower mass galaxies, resulting in stripping of their hot halo gas and initiating strangulation.

\citet{stein16} used ram-pressure stripping simulations employing the moving-mesh code AREPO \citep{springel10} to follow at high resolution the interaction of a galaxy cluster with infalling galaxies. They constructed several different galaxy models, including variations of the amount of gas in the halos of the galaxies, and collided these models with galaxy clusters of different mass. In agreement with our observational results of an \emph{slow-then-rapid quenching} in clusters, \citet{stein16} found that, typically, their model galaxies continue to form stars with only slightly modified rates as a result of the \emph{stripping of the hot gaseous halos} (strangulation) of the galaxies. On the other hand, the cold gas is stripped only 
during pericenter passages with small pericenter distances leading to a full quenching of star formation on a short timescale.

%
%##############################################################################
%######################################################################

\section{Summary}
\label{sec:summary}

This study of environmental effects on the Z(M$_{*}$,SFR) relation is based on Hectospec spectroscopy of cluster galaxies in seven clusters from LoCuSS at $0.15<z<0.26$. The main results can be summarized as follows:

~~~1. Cluster and field galaxies at $0.15<z<0.26$ show, on average, similar MZR and mass-SSFR relations for stellar masses $10 \leq \rm{log(M/M_{\odot})} \leq 11$  (Figs.\,\ref{fig:MassSSFR} and \ref{fig:MassOH}). 
There is, however, tentative evidence ($2.4 \sigma$ significance)
% A significant difference is, however, 
that $10 \leq \rm{log(M/M_{\odot})} < 10.5$ galaxies inside $R_{200}$ 
(red symbols and histograms in Fig.\,\ref{fig:MassOH}) have metallicities displaced to higher values
% and typically lower [OIII]/H$\beta$ ratios (Fig.\,\ref{fig:BPT}) 
compared to field counterparts and to cluster galaxies at $R>R_{200}$.
%Their lower [OIII]/H$\beta$ ratios indicate that these galaxies inside $R_{200}$ are likely galaxies in the act of quenching their star formation according to a recent study of \citet{quai18}.

~~~2. While the measured O/Hs of field and $R>R_{200}$ cluster galaxies are in quite good agreement with FMR predictions, $10 \leq \rm{log(M/M_{\odot})} < 10.5$   cluster galaxies inside $R_{200}$  have a distribution shifted to higher metallicities than predicted by the \citet{lilly13} models with inflowing gas (Fig.\,\ref{fig:DiffLi13}), indicating that a strangulation scenario in which gas inflow has stopped producing enhanced metallicities is a plausible mechanism.
For $10.5 \leq \rm{log(M/M_{\odot})} \leq 11$ galaxies, the MZR saturates and becomes flat for cluster and field galaxies (Fig.\,\ref{fig:MassOH}). Therefore,  strangulation does not have any visible effect on the metallicities of $10.5 \leq \rm{log(M/M_{\odot})} \leq 11$ objects. 

~~~3.  We derive quenching timescales $ \Delta t = 1-2$\,Gyrs, defined as the delay time between the time the galaxies pass $R_{200}$ until \emph{complete} quenching of star formation. This is done by comparing the fraction of star forming cluster galaxies as a function of clustercentric radius to model curves derived from the Millenium simulation (Fig.\,\ref{fig:fSF}). The model curves with the fraction of star forming cluster galaxies as a function of clustercentric radius show the effects of delaying the moment at which the complete quenching occurs, by terminating star formation only in those galaxies accreted into the cluster more than $\Delta t$ Gyrs prior to observations.

The main conclusion is that a \emph{strangulation scenario} in a \emph{slow-then-rapid framework} is plausible to explain the metallicities and specific SFRs of field and cluster galaxies at different clustercentric radii, indicating that the inflow of gas is being stopped when cluster galaxies pass $R_{200}$,
as a result of cluster environmental effects, likely due to the removal of the \emph{hot} halo gas by ram-pressure effects. During the slow-quenching phase of $\Delta t = 1-2$Gyrs these metal-richer galaxies inside $R_{200}$ are still star-forming.
After this $\Delta t$, meanwhile galaxies have travelled to denser inner regions of the cluster where the ram-pressure also exceeds the gravitational restoring pressure of the \emph{cold} gas, their star-formation is likely completely quenched by ram-pressure stripping of their cold gas in the ``rapid'' phase.

%######################################################################

%%%%%%%%%%%%%%%%%%%%%%%%%%%%%%%%%%%%%%%%%%%%%%%%%%%%%%%%%%%%%%%%%%%%

%TBD

%##############################################################################

%\begin{acknowledgements}
%This publication is supported by the Austrian Science Fund (FWF).
%\end{acknowledgements}

%###########################################################################

%\clearpage

%########################################################################

\end{document}